\documentstyle[aps,prb,epsfig,multicol]{revtex}

\begin{document}

\preprint{SNUTP 01-}

\title{Dynamic transitions and resonances in Josephson-junction arrays\\ 
under oscillating magnetic fields}

\author{Gun Sang Jeon}
\address{Center for Strongly Correlated Materials Research,
Seoul National University, Seoul 151-747, Korea}
\author{Hyun Jin Kim and M.Y. Choi}
\address{Department of Physics and Center for Theoretical Physics, 
Seoul National University, Seoul 151-747, Korea}
\author{Beom Jun Kim and Petter Minnhagen}
\address{Department of Theoretical Physics, Ume\aa University, 901 87 Ume\aa,
Sweden}

\maketitle
\draft

\begin{abstract}
We investigate dynamic transitions and stochastic resonance
phenomena in two-dimensional fully frustrated Josephson-junction
arrays driven by staggered oscillating magnetic fields. 
As the temperature is lowered, 
the dynamic order parameter, defined to be the average staggered magnetization, 
is observed to acquire nonzero values. 
The resulting transition is found to belong to the same universality as
the equilibrium $\rm{Z}_2$ transition for small driving amplitudes while 
large driving fields appear to induce deviation from the universality class. 
The transition is also manifested by the stochastic resonance peak
of the signal-to-noise ratio,
which develops above the transition temperature.
\end{abstract}


\begin{multicols}{2}

The two-dimensional (2D) fully frustrated $XY$ model, 
which provides a good description of the
fully frustrated Josephson-junction array (FFJJA), has attracted much
attention for the past decades, with regard to the $\rm{Z}_2$ symmetry 
in addition to the U(1) symmetry present in the system.~\cite{FFXY,double,SLee}
Extensive studies have been devoted to the nature of the equilibrium 
transition in the system, leading to general consensus on double transitions 
with some controversy as to the critical behavior of the Z$_2$
transition.~\cite{double,SLee}
Recently, the kinetic Ising model, which is a prototype model with
the Z$_2$ symmetry, has been examined in the presence of 
oscillating magnetic fields.~\cite{Tome,Acharyya99,Korniss,IsingSR,DoubleSR} 
It has been revealed that spontaneous symmetry breaking takes place for finite 
strength of the oscillating field~\cite{Tome,Acharyya99,Korniss} 
and that in two dimensions the transition belongs to the 
same universality class as the equilibrium 2D Ising
transition.~\cite{Korniss}  
The signal-to-noise ratio (SNR) has also been found to exhibit 
double peaks,~\cite{IsingSR} one above the transition temperature and the
other below the transition temperature, the emergence of which
has been explained in terms of the time-scale matching between the inherent relaxation
time of the system and the external periodic driving.~\cite{DoubleSR} 
Such interesting results in the kinetic Ising model
invokes interest in the dynamics of the FFJJA,
which possesses the same Z$_2$ symmetry as the Ising
model together with the additional continuous U(1) symmetry. 
Furthermore, unlike the Ising model, the FFJJA has real intrinsic 
dynamics derived from the Josephson relations and thus grants direct
experimental realizations;~\cite{review} 
this makes the study of the FFJJA more attractive. 
It is a challenge to prepare oscillating fields which 
couples relevantly to the Z$_2$ degrees of freedom in the FFJJA.
One possible realization may be provided by a periodic array of magnetic 
particles~\cite{Martin} which is placed under the FFJJA and 
driven back and forth periodically.~\cite{DoubleSR}
There have been some studies on the dynamic properties of the
FFJJA,~\cite{Dynamics,Luo} and very recently dynamic transitions 
were investigated in the FFJJA driven by uniform dc current.~\cite{Dominguez}
However, no results are available yet 
about dynamic transitions and stochastic resonance (SR) phenomena
in the presence of oscillating fields.

This paper investigates the dynamic behavior of the FFJJA
in the presence of the transverse magnetic field, which is staggered in space
and periodic in time, with attention paid to the dynamic transition and SR phenomena. 
At low temperatures the chirality in the FFJJA displays antiferromagnetic ordering, 
to which the staggered oscillating field plays the same role 
as the oscillating field in the kinetic Ising model. 
To describe the dynamic transition to the antiferromagnetic ordering, 
we define the dynamic order parameter as the staggered magnetization
averaged over one period of the (staggered) driving field,
and examine its behavior with respect to the amplitude and the frequency of the
driving field as well as the temperature. 
For small driving amplitudes, the scaling plot strongly indicates that 
the transition belongs to the same universality as
the equilibrium $\rm{Z}_2$ transition. 
On the other hand, as the driving amplitude is raised, apparent deviations 
can be observed in the scaling plot, 
suggesting a universality class different from the Z$_2$ one. 
We also examine the SNR and find a peak 
above the transition temperature, disclosing the SR present in the system. 
The absence of the lower peak is discussed in relation to the relaxation time 
of the system.

We begin with the equations of motion
for the phases $\{\phi_i\}$ of the superconducting order 
parameters in superconducting grains forming an $L\times L$ square lattice. 
In the resistively-shunted-junction model 
with the fluctuating twist boundary conditions,~\cite{FTBC}
they read:
\begin{equation} \label{eq:dyn1}
{\sum_j}' \Big[ 
{d \widetilde{\phi}_{ij} \over dt} 
+ \sin (\widetilde{\phi}_{ij}-{\bf r}_{ij} \cdot {\bf \Delta}) 
 + \eta_{ij}\Big]=0,
\end{equation}
where the primed summation runs over the nearest neighbors of grain $i$. 
We have used the abbreviations 
$\widetilde{\phi}_{ij}\equiv \phi_i{-}\phi_j{-}A_{ij}$ and 
${\bf r}_{ij}\equiv {\bf r}_i{-} {\bf r}_j$
with ${\bf r}_i = (x_i , y_i )$ denoting the position of grain $i$, 
and expressed the energy and the time in units of $\hbar I_c/2e$
and $\hbar/2eRI_c$, respectively, with the critical current $I_c$
and the shunt resistance $R$.
The thermal noise current $\eta_{ij}$ is assumed to be white
satisfying
$\langle \eta_{ij}(t{+}\tau) \eta_{kl} (t) \rangle
= 2 T  \delta (\tau) (\delta_{ik}\delta_{jl} {-}
\delta_{il}\delta_{jk})$ 
at temperature $T$ and the dynamics of the twist variables 
${\bf \Delta} \equiv (\Delta_x, \Delta_y )$ is governed by the equations
\begin{equation} \label{eq:dyn2}
\frac{d\Delta_{a}}{dt} 
=\frac{1}{L^2}
\sum_{\langle ij \rangle_{a}}
\left[ \sin(\widetilde{\phi}_{ij}-\Delta_{a})-
{d A_{ij} \over dt} \right]
+ \eta_{\Delta_{a}} , 
\end{equation}
where $L$ is the system size, 
$\sum_{\langle ij \rangle_{a}}$ denotes the
summation over all nearest-neighboring pairs in the $a$-direction $(a = x, y)$, 
and $\eta_{\Delta_a}$ satisfies 
$\langle \eta_{\Delta_a}(t{+}\tau) \eta_{\Delta_a} (t) \rangle
= (2 T / L^2) \delta (\tau)$. 
The gauge field $A_{ij}\, (= - A_{ji})$, which incorporates
the effects of the transverse magnetic field, takes the form
\begin{equation}
A_{ij} = \left\{ 
\begin{array}{ll}
  0 
   & \hbox{for } {\bf r}_j = {\bf r}_i +\hat{\bf x},\\
  \pi x_i + \pi (-1)^{x_i + y_i} f_0\sin \Omega t 
   & \hbox{for } {\bf r}_j = {\bf r}_i + \hat{\bf y},\\
\end{array}
\right. 
\end{equation}
which corresponds to the combined uniform constant and staggered oscillating fields. 
The frustration, i.e., the flux in units of the flux quantum 
through the plaquette at the dual lattice site 
${\bf R}_i \equiv {\bf r}_i + (1/2)(\hat{\bf x}+\hat{\bf y})$
thus consists of the uniform full dc component and the staggered ac component
of amplitude $f_0$ and frequency $\Omega$:
\begin{equation} \label{eq:oscf}
f_{\bf R_i} = {1 \over 2} + (-1)^{x_i + y_i} f_0  \sin \Omega t .
\end{equation}

For the study of the transition associated with the Z$_2$ symmetry
in the FFJJA, it is convenient to consider the chirality
\begin{equation}
C({\bf R},t) \equiv {\rm sgn}\left[\sum_P 
\sin \left(\widetilde{\phi}_{ij}(t) - {\bf
r}_{ij}\cdot {\bf \Delta}(t) \right)\right]
\end{equation}
and the staggered magnetization
\begin{equation}
m(t) \equiv {1 \over L^2 } \sum_{\bf R} (-1)^{x_i + y_i} C({\bf R},t),
\end{equation}
where $\sum_P$ denotes the directional plaquette
summation of links around dual lattice site $\bf R$. 
To probe the dynamic transition in the presence of an oscillating field, 
we define the dynamic order parameter as the staggered magnetization
averaged over the period of the field~\cite{Tome}
\begin{equation}
Q = {\Omega \over 2\pi} \left| \oint m(t) dt \right| .
\end{equation}
In numerical simulations, the sets of equations of motion (\ref{eq:dyn1}) and
(\ref{eq:dyn2}) have been integrated via the modified Euler method 
with time steps $\Delta t = 0.05$. 
We have followed an annealing schedule with the equilibration time at
least 500 periods at each temperature.

\begin{figure}
\centerline{\epsfig{file=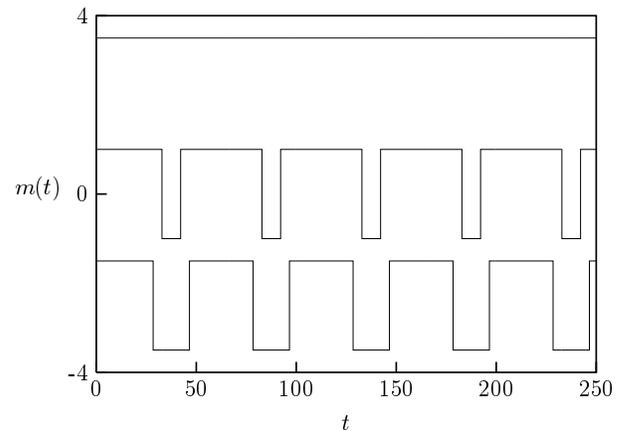,width=8cm}}
\vspace{0.2cm}
\caption{Time evolution of the staggered magnetization at zero temperature 
for driving frequency $\Omega/2\pi = 0.02$ and 
amplitudes $f_0 = 0.4, 0.6$, and $1.2$ from above.
For clarity, the data corresponding to $f_0 =0.4$ and $1.2$ have been
shifted by 1.5 upward and downward, respectively.}
\label{fig:zeroTm}
\end{figure}

\begin{figure}
\centerline{\epsfig{file=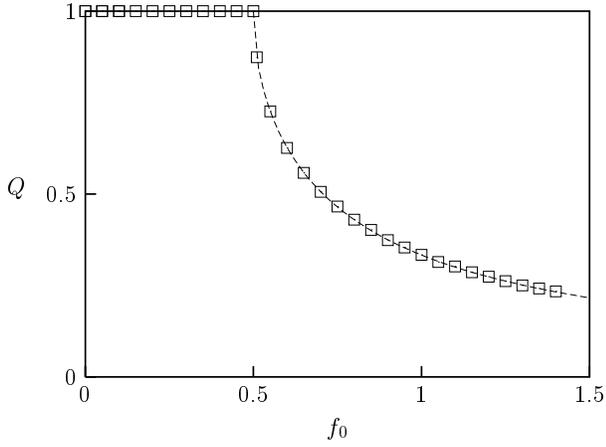,width=8cm}}
\vspace{0.2cm}
\caption{Dynamic order parameter as a function of $f_0$ at zero temperature 
in the system of size $L=16$ and with frequency $\Omega/2\pi = 0.02$. 
The dashed line represents the analytic result 
in Eq.~(\ref{eq:Qf0}).}
\label{fig:zeroTQ}
\end{figure}

In the fully frustrated array, it is well known that its Z$_2$
symmetry produces two kinds of antiferromagnetic chirality ordering at
zero temperature.
The staggered driving field in Eq.~(\ref{eq:oscf}), which favors
one of the two ground states in the first half of a period and the other in
the latter half, is expected to induce oscillations between the two 
ground states when the driving amplitude is sufficiently large.
Figure~\ref{fig:zeroTm} displays the time evolution of the 
staggered magnetization $m(t)$ at zero temperature, 
with the initial condition $m(t{=}0)=1$, 
for various amplitudes of the driving field.
For $f_0 < f_c$\,(which is equal to 0.5 regardless of the driving frequency),
the system remains in the state with $m=1$
even in the presence of the driving field. 
As the amplitude is increased beyond the critical value $f_c$, 
the system is driven out of the $m=1$ state, giving rise to
oscillations in the staggered magnetization between $m=\pm 1$.
It is interesting that the residence time in one state is different
from that in the other.  Accordingly, for $f_0 > f_c$, 
although the dynamic order parameter is not unity, it still does not vanish. 
>From Eq.~(\ref{eq:oscf}), we expect the array to stay at the $m=-1$ state 
only during the interval given by $f_0 \sin \Omega t < -1/2$; 
this leads to the dynamic order parameter for $f_0 > f_c$,
\begin{equation} \label{eq:Qf0}
Q(f_0) = 1 - {2 \over \pi} \cos^{-1} {1 \over 2 f_0 }, 
\end{equation}
which depends on the driving amplitude $f_0$ but not on the driving frequency
$\Omega$. 
In Fig.~\ref{fig:zeroTQ} we plot the dynamic order parameter $Q$ versus 
the amplitude $f_0$ at zero temperature. 
The numerical data indeed agree very well with
the analytical results given by Eq.~(\ref{eq:Qf0}), 
establishing the validity of the analytical argument. 

\begin{figure}
\centerline{\epsfig{file=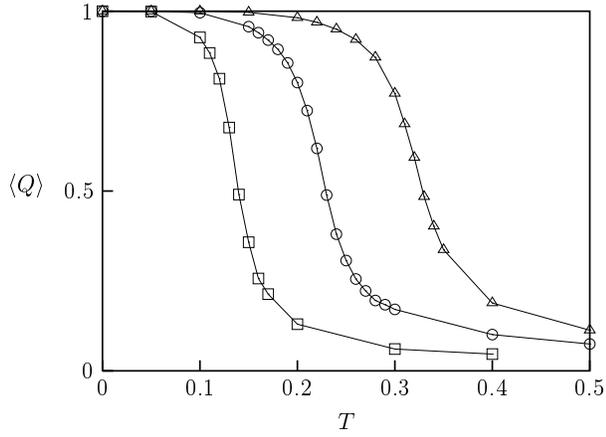,width=8cm}}
\vspace{0.2cm}
\centerline{(a)}
\vspace{0.2cm}
\centerline{\epsfig{file=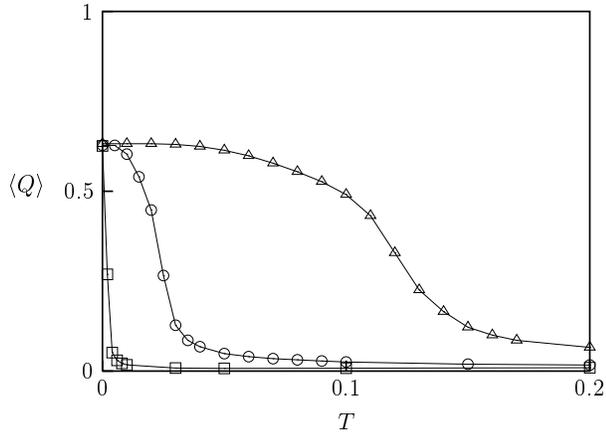,width=8cm}}
\vspace{0.2cm}
\centerline{(b)}
\vspace{0.2cm}
\caption{Dynamic order parameter as a function of temperature $T$ in the system 
of size $L=16$ for various driving frequencies 
$\Omega/2\pi = 0.02(\Box)$, $0.04 (\bigcirc)$, $0.08 (\triangle)$ and 
(a) $f_0=0.3$; (b) $f_0=0.6$ .}
\label{fig:Q}
\end{figure}

At sufficiently high temperatures, the influence of the oscillating field 
may be neglected and the staggered magnetization fluctuates randomly with time,
resulting in the null value of the dynamic order parameter.
Accordingly, we expect that there is a transition between the ordered state at low
temperatures and the disordered state at high temperatures. 
Figure~\ref{fig:Q}, in which the behavior of $\langle Q\rangle$, the ensemble average 
of the dynamic order parameter, with the temperature is plotted 
for various driving frequencies, 
demonstrates clearly the existence of such a transition: 
The dynamic order parameter, starting from zero at high temperatures, 
develops as the temperature is reduced.  
It then grows rapidly around a certain temperature, from which the 
transition temperature $T_c$ can be inferred, 
and saturates eventually as the temperature approaches zero. 
The saturation value assumed by the dynamic order parameter at zero temperature 
is unity for $f_0 <f_c$, as shown in Fig.~\ref{fig:Q}(a); 
for $f_0 >f_c$, it is given by Eq.~(\ref{eq:Qf0}) [see Fig.~\ref{fig:Q}(b)]. 
It is indicated that in both cases the transition temperature becomes higher with the
increase of the driving frequency $\Omega$.

\begin{figure}
\centerline{\epsfig{file=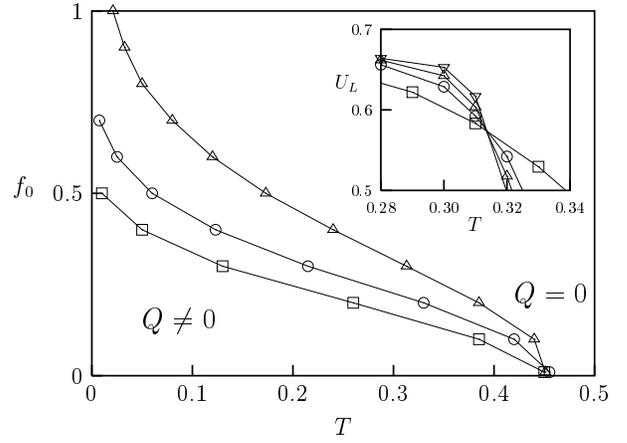,width=8cm}}
\vspace{0.2cm}
\caption{Dynamic phase diagram on the $T{-}f_0$ plane for various driving 
frequencies $\Omega/2\pi = 0.02(\Box)$, $0.04(\bigcirc)$, $0.08(\triangle)$. 
The boundaries are determined by the crossing points of Binder's
cumulant for several sizes $L=8, 16, 24$, and $32$. 
Inset: Binder's cumulant $U_L$ as a function of the temperature for $f_0 =0.3$ 
and $\Omega/2\pi = 0.08$ in the system of size
$L=8(\Box)$, $16(\bigcirc)$, $24(\triangle)$, and $32(\bigtriangledown)$.
}
\label{fig:PD}
\end{figure}

The transition temperature in the thermodynamic limit can be conveniently
obtained from the consideration of Binder's cumulant~\cite{Binder}
\begin{equation}
U_L = 1 - {\langle Q^4\rangle \over 3 \langle Q^2 \rangle^2 }. 
\end{equation}
This quantity is known to be independent of the system size at the transition 
temperature $T_c$, which allows one to estimate the transition temperature
by the crossing point of the cumulants for different sizes.  
[See the inset of Fig.~\ref{fig:PD}, in which the cumulant is plotted as 
a function of the temperature $T$ for various system sizes.] 
In this manner we have estimated the transition temperature 
for various values of the driving amplitude $f_0$ and frequency $\Omega$,
and show in Fig.~\ref{fig:PD} the resulting phase diagram for the dynamic transition 
on the $T{-}f_0$ plane. 
For given frequency $\Omega$, the transition temperature decreases 
monotonically with the driving amplitude $f_0$, 
similarly to the case of the kinetic Ising model. 
As $\Omega$ is increased, on the other hand, the transition temperature 
also increases, expanding the ordered region on the $T{-}f_0$ plane. 
It is thus concluded that high-frequency driving helps to establish order
whereas large-amplitude one tends to suppress order. 
It should be noted here that the cooling and heating curves
for the dynamic order parameter do not exhibit any
hysteresis in the strong-field regime. 
This apparently indicates the absence of a tricritical point, 
above which the system undergoes a discontinuous transition, 
and contrasts sharply with the kinetic Ising models under oscillating 
fields both in the mean-field version~\cite{Tome} and in two
dimensions.~\cite{Acharyya99}

\begin{figure}
\centerline{\epsfig{file=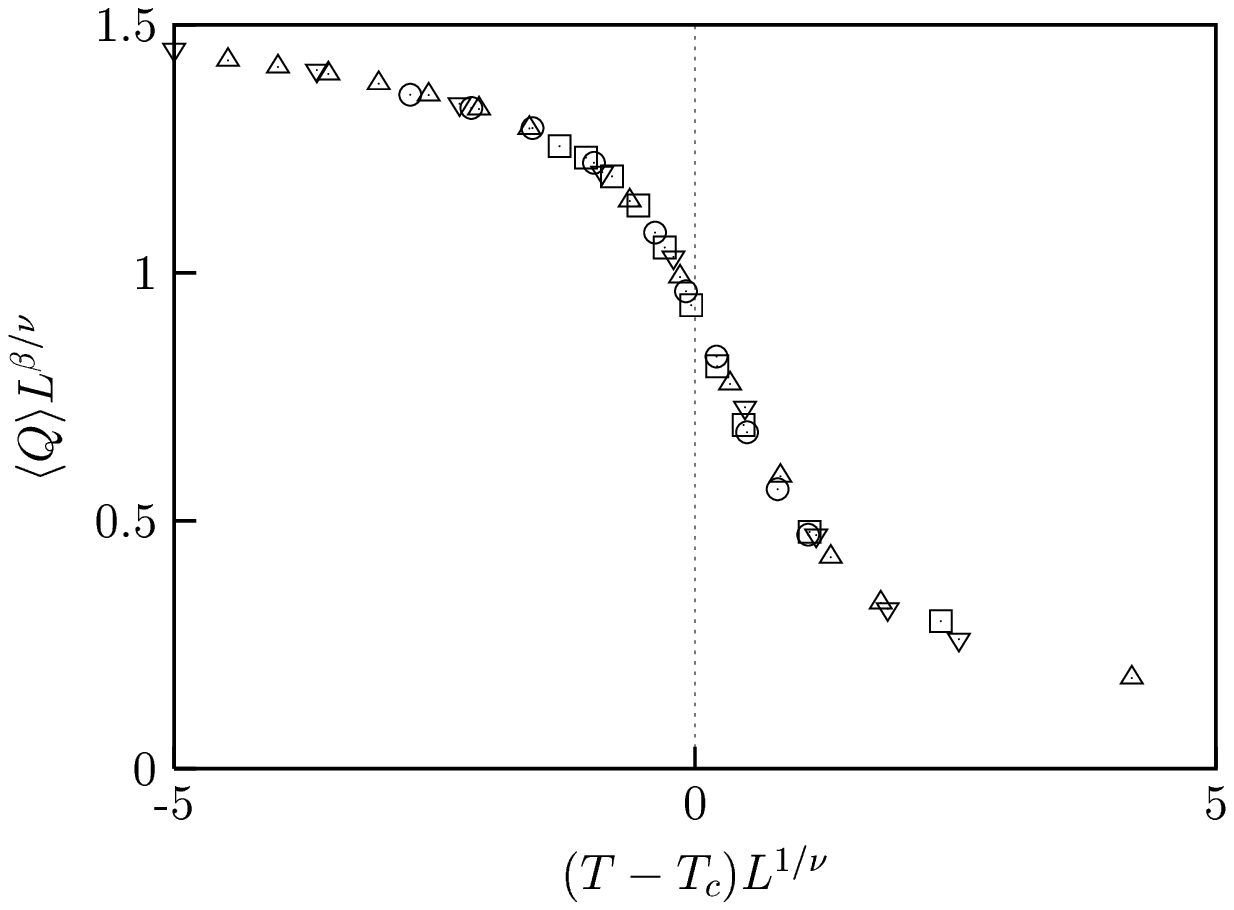,width=8cm}}
\vspace{0.2cm}
\centerline{(a)}
\vspace{0.2cm}
\centerline{\epsfig{file=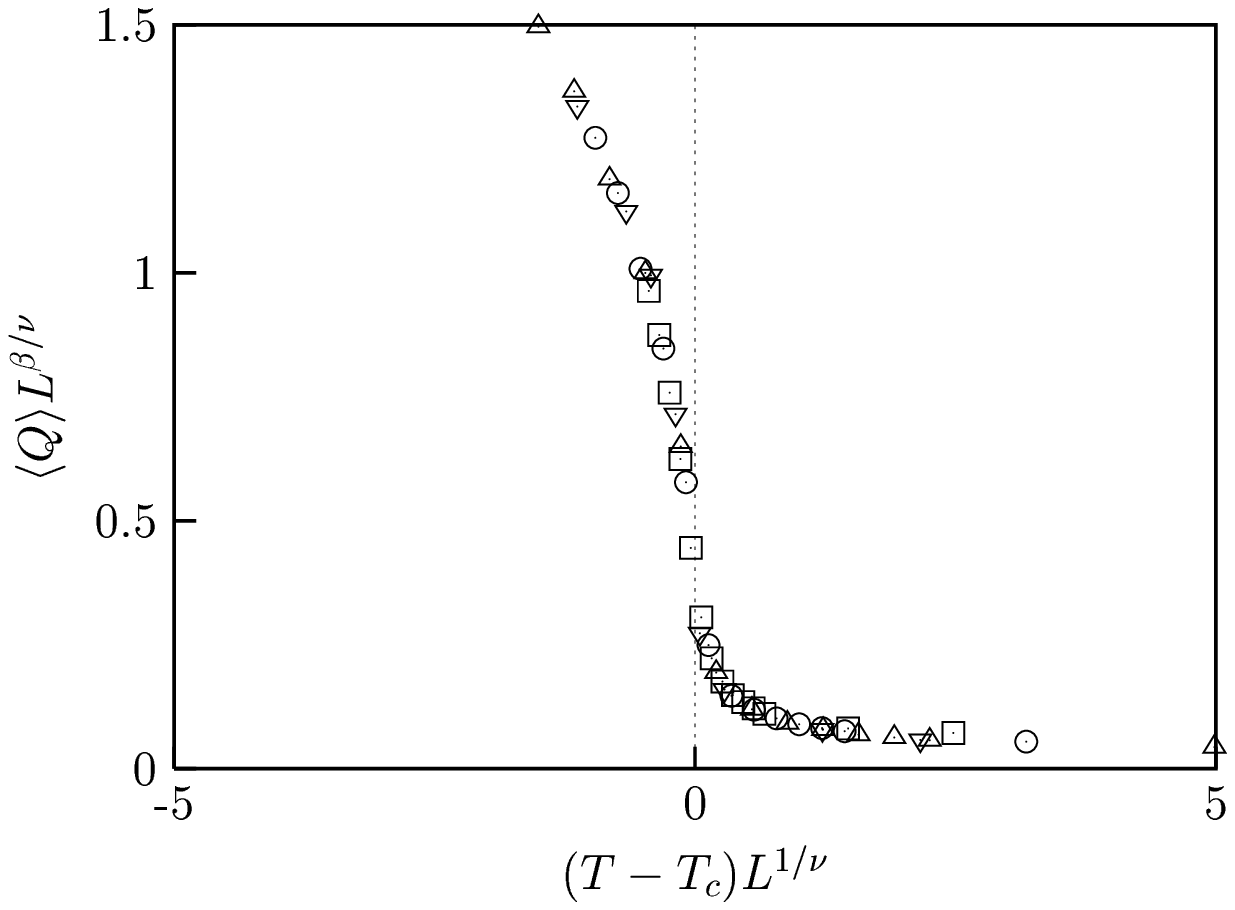,width=8cm}}
\vspace{0.2cm}
\centerline{(b)}
\vspace{0.2cm}
\caption{Scaling plot of the dynamic order parameter versus the temperature 
for size $L=8(\Box)$, $16(\bigcirc)$, $24(\triangle)$, and $32(\bigtriangledown)$ 
with $\Omega/2\pi=0.08$ and (a) $f_0=0.3$; (b) $f_0=0.8$. 
The best collapse is obtained with (a) $\beta=0.10$ and $\nu=0.82$; 
(b) $\beta=0.36$ and $\nu=0.90$. 
}
\label{fig:scaling}
\end{figure}

Regarding the nature of the transition, it is revealing to recall that 
the system at zero temperature, residing in one of the two ($m=\pm 1$) states
for $f_0 < f_c$, displays oscillations between the two states
as $f_0$ is raised beyond $f_c$. 
This suggests that the critical behavior near the transition may change 
qualitatively at $f_0 =f_c$.
To probe such possibility, 
we consider the scaling law for the dynamic order parameter~\cite{FSS}
\begin{equation} \label{eq:scaling}
\langle Q \rangle  = L^{-\beta/\nu} F((T{-}T_c) L^{1/\nu})
\end{equation}
with appropriate scaling function $F(x)$
and plot $\langle Q \rangle  L^{\beta/\nu}$ versus $(T-T_c) L^{1/\nu}$ 
in Fig.~\ref{fig:scaling} for various sizes.
Figure~\ref{fig:scaling}(a) shows the best collapse of
the numerical data for $f_0=0.3$, yielding the critical exponents 
$\beta=0.10\pm 0.02$ and $\nu=0.82\pm 0.05$; 
these values provide good collapse in the whole range of $f_0 < f_c$. 
Note the good agreement of the obtained exponents with those
of the equilibrium transition~\cite{SLee}, which 
strongly suggests that the dynamic transition at weak driving fields 
belongs to the same universality class as the equilibrium Z$_2$ transition 
in the fully frustrated $XY$ model. 
Similar conclusion was also reached in the kinetic Ising model under 
oscillating fields.~\cite{Korniss}
For $f>f_c$, on the other hand, the scaling plot with the same
critical exponents, apparently does not collapse into a single curve. 
Such dispersion, which begins to show up around $f_0 = f_c$, 
indicates deviation of the critical behavior from that of 
the equilibrium Z$_2$ transition, 
manifesting the expected change of the nature of the transition at $f_0 =f_c$.
Instead the data fit well to the scaling form in Eq.~(\ref{eq:scaling})
with $\beta=0.36\pm0.08$ and $\nu=0.9\pm 0.1$, 
as shown in Fig.~\ref{fig:scaling}(b) for $f_0 = 0.8$.
These values of the exponents, estimated from the best collapse for $f_0 > f_c$, 
turn out to be independent of $f_0$ within numerical errors.

\begin{figure}
\centerline{\epsfig{file=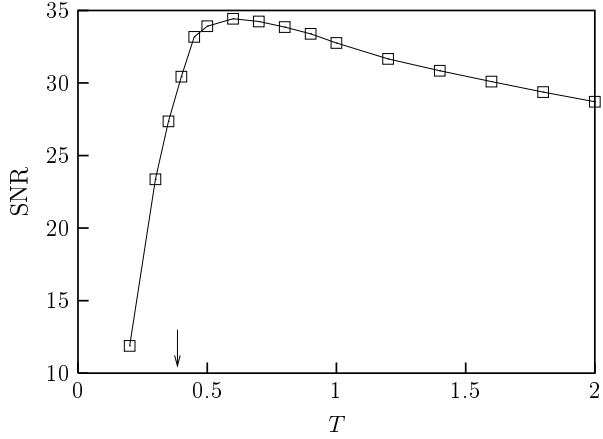,width=8cm}}
\vspace{0.2cm}
\centerline{(a)}
\vspace{0.2cm}
\centerline{\epsfig{file=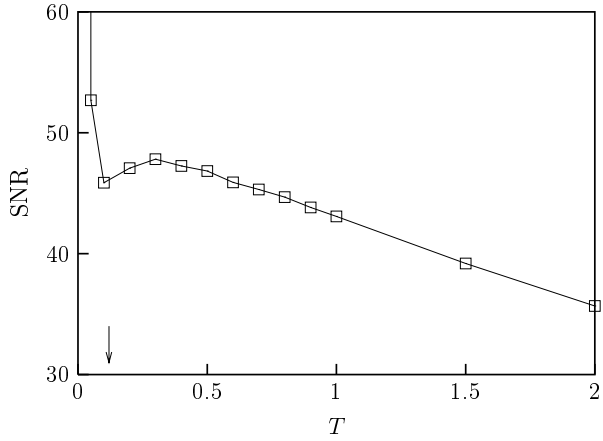,width=8cm}}
\vspace{0.2cm}
\centerline{(b)}
\vspace{0.2cm}
\caption{Signal-to-noise ratio versus the temperature in the system of size
$L=16$ for (a) $f_0=0.1$ and $\Omega/2\pi=0.02$; (b) $f_0=0.6$ and
$\Omega/2\pi=0.08$. The arrows indicate the positions of the transition 
temperatures.}
\label{fig:SNR}
\end{figure}

Finally, to explore the possibility of the SR phenomena in the system, 
we compute the SNR defined to be 
\begin{equation}
{\rm SNR} = 10 \log_{10} \left[ {S \over N} \right]. 
\end{equation}
Here the signal $S$ is given by the peak value of the power spectrum of
$m(t)$ at the driving frequency $\Omega$ while the background noise level $N$ 
is estimated by the average power spectrum around the signal peak. 
Figure~\ref{fig:SNR} displays the SNR versus the temperature (a) for 
weak driving $(f_0 = 0.1)$ and (b) for strong driving $(f_0 = 0.6)$. 
For $f_0 < f_c$ shown in (a), the SNR peak emerges above the transition temperature 
and signals the presence of the SR in the system, 
which is reminiscent of the SR behavior of the kinetic Ising field in weak
oscillating fields.~\cite{IsingSR} 
For $f_0 > f_c$ one may expect monotonic decrease of the SNR with the
temperature since the array has inherent oscillations of the staggered 
magnetization $m(t)$ even at zero temperature. 
This expectation appears to be valid at low temperatures as shown in 
Fig.~\ref{fig:SNR}(b). 
As the temperature is raised, however, the SNR changes its behavior: 
Sharp decrease of the SNR turns into gradual increase 
seemingly at the transition temperature and into gradual decrease again 
above the transition temperature, giving rise to a small broad peak. 

\begin{figure}
\centerline{\epsfig{file=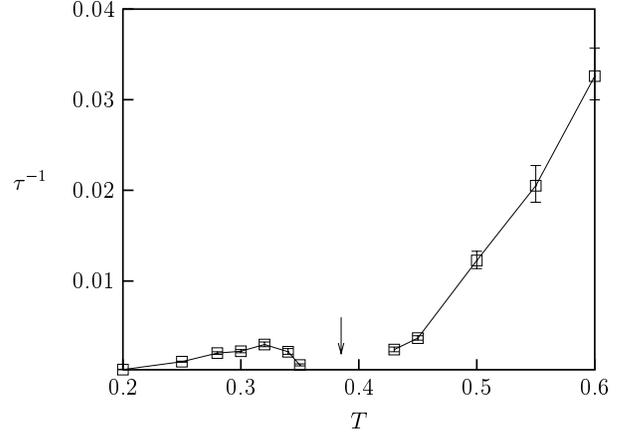,width=8cm}}
\vspace{0.2cm}
\caption{Inverse of the relaxation time as a function of the temperature for
$L=16$, $f_0=0.1$, and $\Omega/2\pi=0.02$.
The arrow indicates the position of the transition temperature.}
\label{fig:inversetau}
\end{figure}

Here it is of interest that the system exhibits no peak below the dynamic
transition,
which seems to be in contradiction to the prediction of the double SR
peaks, one below and the other above the dynamic transition
temperature.~\cite{DoubleSR}  
The origin of the double peaks is the time-scale matching between the
external periodic driving and the relaxation time associated with the
transition in the system.
To disclose the origin of the discrepancy, we examine the relaxation
of the system from random initial configurations at given temperatures,
and estimate the relaxation time $\tau$ by fitting 
$Q(t)-\langle Q \rangle$ to the exponential form 
$\sim \exp [ - t / \tau]$ in the long-time scale. 
The inverse of the resulting relaxation time for $f_0 =0.1$ is shown in 
Fig.~\ref{fig:inversetau}. 
As expected, the inverse relaxation time increases with the
temperature above $T_c$, 
which is consistent with the critical behavior 
\begin{equation}
\tau^{-1} \sim (T-T_c)^{\nu z}
\end{equation}
with the critical exponents given by the equilibrium values
$\nu=0.813\pm 0.005$ and $z=2.17\pm 0.04$.~\cite{Luo} 
Thus supported again is the proposition that the dynamic transition under weak 
driving fields belongs to the same universality class as the equilibrium transition.
At low temperatures below $T_c$, on the other hand, 
it is observed that the relaxation time is enhanced enormously.
Due to such enhancement of the relaxation time below $T_c$, the system 
fails to match the external time scale, which is estimated to 
be $\tau^{-1} \approx 0.032$ from the SNR peak above $T_c$. 
The origin of the enhancement at low temperatures is not clear yet, but
it can be attributed to the existence of a large number of metastable states
associated with, e.g., domain-wall excitations, which cannot be observed in the
mean-field system considered in Ref.~\onlinecite{DoubleSR}. 
Indeed faceted domain walls have been pointed out to cause slow growth 
in the FFJJA below $T_R \approx 0.3$,~\cite{SJLee} where
the enhancement of the relaxation-time data is conspicuous 
in Fig.~\ref{fig:inversetau}. 
This strongly supports our argument concerning the failure of time-scale 
matching.

In summary, we have studied dynamic properties of the 2D FFJJA driven by 
staggered oscillating magnetic fields, 
paying attention to the dynamic transition and SR phenomena.  
To describe the dynamic transition to the antiferromagnetic ordering 
of the chirality in the system, 
we have introduced the dynamic order parameter, which is the staggered magnetization
averaged over one period of the staggered driving field,
and examined its behavior with respect to the amplitude and the frequency of the
driving field as well as the temperature. 
The transition temperature has been computed by means of Binder's cumulants 
for various driving amplitudes and frequencies, and as a result,
the phase diagram displaying boundaries of the dynamic transition has been 
obtained on the plane of the temperature and the driving amplitude 
for various driving frequencies.
While the scaling analysis for weak driving has demonstrated the universality of
the equilibrium $\rm{Z}_2$ transition, 
a universality class different from the Z$_2$ one has been suggested for 
strong driving. 
We have also examined the SNR and found a peak 
above the transition temperature, disclosing the SR present in the system. 
The absence of the lower peak has been discussed in relation to the relaxation time 
of the system and attributed to the striking increase at low temperatures. 

This work was supported in part by the Ministry of Education of Korea through the BK21 
Program (H.J.K and M.Y.C.) and by the Swedish Natural Research Council through 
Contract No. F 5102-659/2001 (B.J.K. and P.M.).

\end{multicols}

\end{document}